\documentclass[twocolumn]{revtex4-2}
\usepackage{graphicx,amsmath, amsfonts,bm, color}

\usepackage{xcolor,hyperref}
\hypersetup{
   colorlinks,
   linkcolor={blue!50!black},
   citecolor={blue!50!black},
   urlcolor={blue!80!black}
}

\renewcommand{\=}{\!=\!}

\DeclareMathOperator{\sgn}{sgn}

\begin{document}

\title{Unconventional singularities, scale separation and energy balance in frictional rupture}
\author{Efim A.~Brener$^{1,2}$}
\author{Eran Bouchbinder$^{3}$}
\email{Corresponding author: eran.bouchbinder@weizmann.ac.il}
\affiliation{$^{1}$Peter Gr\"unberg Institut, Forschungszentrum J\"ulich, D-52425 J\"ulich, Germany\\
$^{2}$Institute for Energy and Climate Research, Forschungszentrum J\"ulich, D-52425 J\"ulich, Germany\\
$^{3}$Chemical and Biological Physics Department, Weizmann Institute of Science, Rehovot 7610001, Israel}

\begin{abstract}
A widespread framework for understanding frictional rupture, such as earthquakes along geological faults, invokes an analogy to ordinary cracks. A distinct feature of ordinary cracks is that their near edge fields are characterized by a square root singularity, which is intimately related to the existence of strict dissipation-related lengthscale separation and edge-localized energy balance. Yet, the interrelations between the singularity order, lengthscale separation and edge-localized energy balance in frictional rupture are not fully understood, even in physical situations in which the conventional square root singularity remains approximately valid. Here we develop a macroscopic theory that shows that the generic rate-dependent nature of friction leads to deviations from the conventional singularity, and that even if this deviation is small, significant  non-edge-localized rupture-related dissipation emerges. The physical origin of the latter, which is predicted to vanish identically in the crack analogy, is the breakdown of scale separation that leads an accumulated spatially-extended dissipation, involving macroscopic scales. The non-edge-localized rupture-related dissipation is also predicted to be position dependent. The theoretical predictions are quantitatively supported by available numerical results, and their possible implications for earthquake physics are discussed.
\end{abstract}

\maketitle

\noindent{\bf \large Introduction}

The failure of frictional systems, composed of bodies interacting along contact interfaces, is mediated by the propagation of interfacial frictional rupture~\cite{Svetlizky2019,Ben-Zion2001,Scholz2002}. A prominent example for such spatiotemporal frictional rupture processes is earthquakes along geological faults~\cite{Scholz2002,Rice1980a,Kostrov1988,Kanamori2004,Ohnaka2013}. A widespread framework for understanding frictional rupture invokes a close analogy to ordinary cracks~\cite{Ida1972,Palmer1973,Kanamori2000,Das2003,Abercrombie2005,Lu2010,Lu2010a,Noda2013a,Svetlizky2014,Kammer2015,Svetlizky2016,Bizzarri2016,Rubino2017,Svetlizky2017a,PartI,Barras2020, madariaga1998modeling, Madariaga2011, peyrat2004nonlinear, herrera2017dynamic, gallovivc2020earthquake}, despite notable differences in the underlying physics. Most importantly, an ordinary tensile (opening) crack that propagates inside a material loaded externally leaves behind it fully-broken, stress-free surfaces, while the interface left behind a frictional rupture front remains in contact and hence features a finite frictional stress (strength) $\tau$. Moreover, not only $\tau$ does not vanish as in ordinary tensile cracks, but in fact it is a dynamical field that varies in space and time, and that is self-selected by the failure dynamics; it depends on the local slip rate/velocity $v$ and on the instantaneous structural state of the frictional interface.
\begin{figure}[h]
\centering
\includegraphics[width=0.45\textwidth]{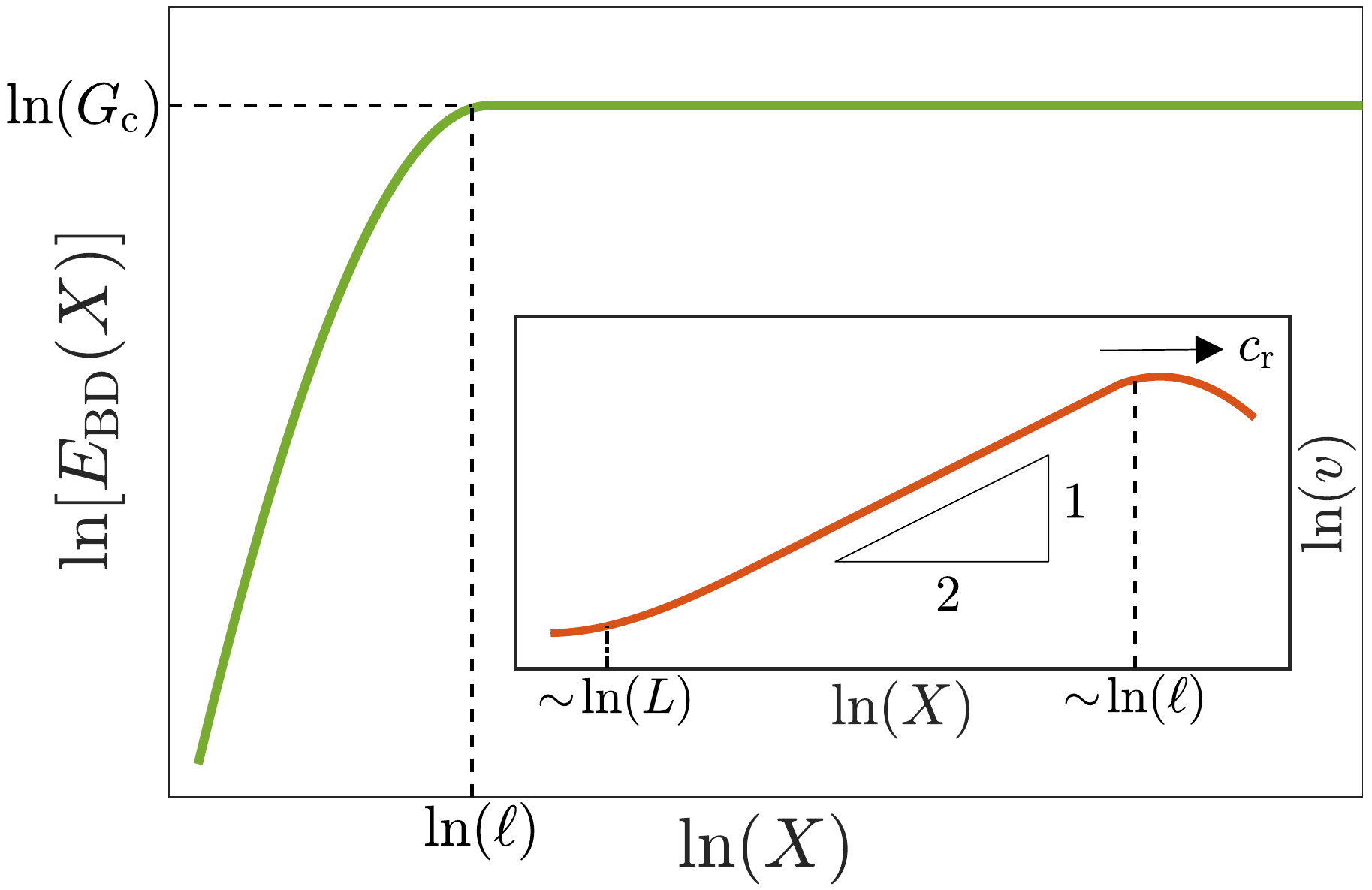}
\caption{{\bf Scale separation and singularity order in conventional shear cracks}. A schematic illustration of dissipation-related lengthscale separation and singularity order in conventional shear cracks. The breakdown energy $E_{\rm BD}$ is plotted as a function of the distance $X$ behind a propagating shear crack (shown in the inset), in logarithmic scale. The breakdown energy $E_{\rm BD}$ quantifies the energy dissipation associated with crack propagation and is given by  $E_{\rm BD}(X)\!=\!G_{\rm c}(\Delta{\cal E}_{_{\rm BD}}(X)+1)$, where the spatial integral $\Delta{\cal E}_{_{\rm BD}}(X)$ is given in Eq.~\eqref{eq:DE} and $G_{\rm c}$ is the fracture energy (defined in the text). $E_{\rm BD}(X)$ increases over the spatial range $0\!<\!X\!<\!\ell$ (i.e.~inside the so-called cohesive/process zone~\cite{Freund1998,Broberg1999}), but saturates at $E_{\rm BD}(X)\!=\!G_{\rm c}$ for $X\!\ge\!\ell$ (see dashed lines). That is, conventional cracks feature strict scale separation, where dissipation occurs only on a localization length $\ell$. (inset) A schematic illustration of the slip velocity field $v(X)$ behind a shear crack propagating at an instantaneous velocity $\dot{L}\!=\!c_{\rm r}$ from left to right ($L$ is the crack length and the dot stands for a time derivative. In addition, note that $X$ here is increasing from right to left, unlike the main panel). $v(X)$ features the conventional Linear Elastic Fracture Mechanics (LEFM) $-\tfrac{1}{2}$ singularity (see triangle and note the logarithmic scale) on intermediate scales, $\ell\!\ll\!X\!\ll\!L$. This square root singularity is directly related to the strict dissipation-related lengthscale separation illustrated in the main panel.}
\label{fig:fig1}
\end{figure}

The rather well-developed theory of ordinary cracks, the so-called Linear Elastic Fracture Mechanics (LEFM)~\cite{Freund1998,Broberg1999}, offers powerful tools that would be very useful for understanding, interpreting and quantifying frictional rupture, if the analogy holds. LEFM is based on scale separation between edge-localized dissipation, which takes place on a short lengthscale $\ell$, and linear elastic driving energy, which is stored on significantly larger scales, larger than the crack length $L$. In particular, cracks in the LEFM framework are characterized by
edge-localized energy dissipation per unit area $G_{\rm c}$ (the so-called fracture energy), which is balanced by an elastic energy flux $G$ into the edge region. The latter is transported from large to small scales by singular fields that are characterized by a universal $-\tfrac{1}{2}$ exponent~\cite{Freund1998,Broberg1999}, valid at intermediate scales between $\ell$ and $L$. These relations between the singularity order, lengthscale separation and edge-localized energy balance are illustrated in Fig.~\ref{fig:fig1}.

In relating frictional rupture to LEFM, one should consider the residual stress $\tau_{\rm res}$, which is finite for frictional rupture, but vanishes for ordinary tensile cracks propagating inside bulk materials under external loading. $\tau_{\rm res}$ is not an intrinsic interfacial quantity, but rather it is an emergent property that is self-selected by the dynamics of the system, through a coupling between the interfacial constitutive relation and bulk elastodynamics~\cite{PartI}. Once $\tau_{\rm res}$ is known, frictional rupture dynamics are quantified relative to a sliding state characterized by $\tau_{\rm res}$. In particular, frictional rupture is then described by the difference between the frictional stress $\tau$ and $\tau_{\rm res}$, i.e.~by $\tau-\tau_{\rm res}$, as will become evident below.

While it is known that LEFM cannot be strictly valid for frictional rupture, where the frictional strength $\tau$ is self-selected and generally depends on the structural state of the frictional interface and on the slip velocity $v$, the conventional LEFM $-\tfrac{1}{2}$ singularity, lengthscale separation and edge-localized energy balance are extensively used in the context of modeling efforts, laboratory experiments and field observations~\cite{Ida1972,Palmer1973,Kanamori2000,Das2003,Abercrombie2005,Lu2010,Lu2010a,Noda2013a,Svetlizky2014,Kammer2015,Svetlizky2016,Bizzarri2016,Rubino2017,Svetlizky2017a,PartI,Barras2020,
madariaga1998modeling, Madariaga2011, peyrat2004nonlinear, herrera2017dynamic, gallovivc2020earthquake}. Yet, to the best of our knowledge, the range of validity of the approximated LEFM picture for frictional rupture, and the interrelations between the singularity order, lengthscale separation and edge-localized energy balance in frictional systems are still not fully understood. Our goal in this paper is to shed basic light on these fundamental issues by developing a comprehensive theory of rupture-related dissipation, lengthscale separation and the singularity order of near rupture edge fields.

We show that the generic rate-dependent nature of friction leads to deviations from the conventional LEFM singularity, and that these deviations can be small if a properly identified dimensionless group of physical parameters is small. We also show that the emergence of unconventional singularities in frictional rupture is accompanied by the breakdown of scale separation, which leads to spatially-extended dissipation that involves macroscopic scales. We show that when the deviation of the unconventional singularity order from the conventional LEFM $-\tfrac{1}{2}$ one is small, edge-localized dissipation $G_{\rm c}$ can be identified on a length $\ell$, but $G_{\rm c}$ can be significantly smaller than rupture-related dissipation (while they are identical in LEFM, cf.~Fig.~\ref{fig:fig1}). Furthermore, the latter is shown to be position dependent. The theory is quantitatively supported by extensive numerical simulations of rate-and-state dependent frictional dynamics, including explaining recent puzzling observations~\cite{Barras2020}. Finally, some possible implications for earthquake physics are discussed.\\

\noindent{\bf \large Results}

In order to study rupture-related dissipation and the associated scales involved, we consider the breakdown energy at an
observation point $x_i$ along the rupture plane, defined as $E_{_{\rm BD}}(t; x_i)\=\int_0^{\delta(\!t;\,x_i\!)}[\tau(\delta'; x_i)-\tau_{\rm res}]\,d\delta'$~\cite{Bizzarri2010a}. Here $x_i$ is a fixed position away from the hypocenter (the nucleation site of frictional rupture, whose instantaneous size is $L(t)$, and nucleation occurred at $t\=0$), $\tau(\delta; x_i)$ the frictional stress at that position and the slip displacement is $\delta(t;x_i)\=\int_{t_{x_i}}^t\!\!v(t; x_i)\,dt$, where $t_{x_i}$ is defined such that $L(t_{x_i})\=x_i$. The term ``breakdown'' refers here to the fact that $E_{_{\rm BD}}$ involves stresses surpassing $\tau_{\rm res}$, i.e.~it does not account for the background frictional dissipation (heat) associated with sliding against the residual stress $\tau_{\rm res}$. Consequently, $E_{_{\rm BD}}$ is the rupture-related dissipation. For ordinary cracks, $E_{_{\rm BD}}(t; x_i)$ is predicted to be independent of $x_i$ and to increase over a short timescale $\ell/c_{\rm r}(t_{x_i})$ (for $t\!>\!t_{x_i}$, where $c_{\rm r}(t)\!\equiv\!\dot{L}(t)$ is the instantaneous crack propagation velocity), until it saturates at $G_{\rm c}$, as illustrated in Fig.~\ref{fig:fig1}. Our first goal is to develop a theory of the breakdown energy $E_{_{\rm BD}}$ for generic rate-and-state frictional interfaces.\\

\begin{figure}[ht!]
\centering
\includegraphics[width=0.45\textwidth]{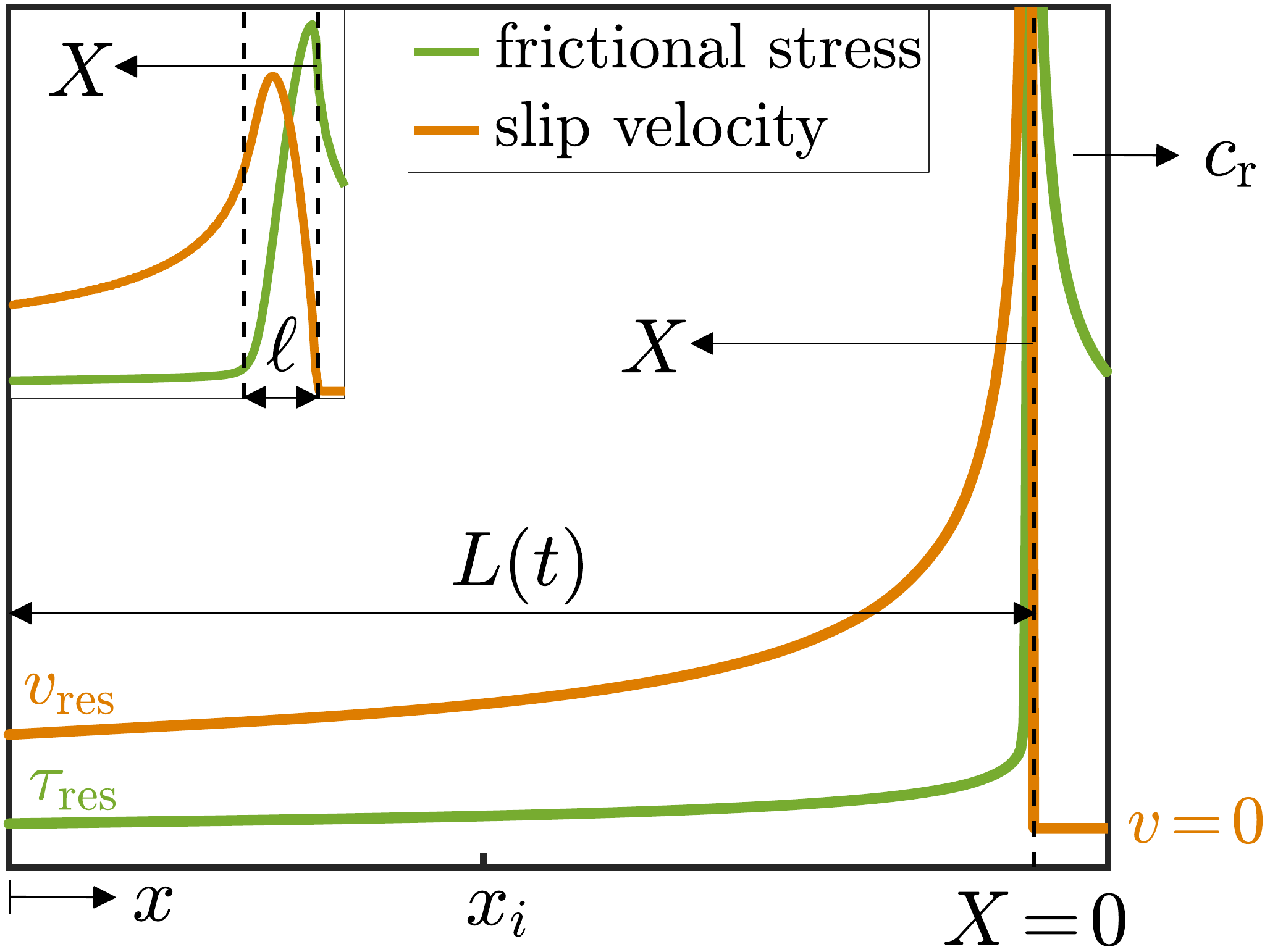}
\caption{{\bf The near-edge fields of a rupture front propagating along a rate-and-state frictional interface}. A frictional rupture front that nucleated at $x\!=\!0$ and propagates to the right in a 2D anti-plane rate-and-state friction simulation (its symmetric counterpart,
propagating to the left, is shown in Fig.~3a of~\cite{Barras2020}, see details about the computer simulation therein). Its instantaneous half-length is $L(t)$ and instantaneous propagation
velocity is $c_{\rm r}(t)\!\simeq\!0.94c_s$. Shown are the frictional stress (strength) field $\tau(X; c_{\rm r},L)$ (green) and slip velocity field $v(X; c_{\rm r},L)$ (orange) left behind the propagating edge.
Here $X(t)$ is a coordinate moving with the edge and pointing backwards (cf.~Fig.~\ref{fig:fig1}), whose origin ($X(t)\!=\!0$) is defined according to $v\!=\!0$. The two fields approach finite residual
values, $\tau_{\rm res}$ and $v_{\rm res}$, respectively, far behind the propagating edge. (inset) A zoom in on the edge region, revealing a localization lengthscale $\ell$
associated with edge-localized dissipation, resulting in an effective fracture energy $G_{\rm c}$ (see Fig.~\ref{fig:fig3}a, for a precise definition of $\ell$). $v(x,t)$ follows, to a very good approximation, the conventional square root singularity of Linear Elastic Fracture Mechanics (LEFM) at an intermediate region, i.e.~$X\!>\!\ell$ and prior to approaching $v_{\rm res}$ (fit not shown here, see~\cite{Barras2020}). The same conventional singularity is featured by $\tau(X,t)$ ahead of the edge, $X\!<\!0$ and $|X|\!>\!\ell$, but it is not discussed here~\cite{Barras2020}.}
\label{fig:fig2}
\end{figure}

\noindent{\bf Theory of the breakdown energy for rate-and-state frictional interfaces.} We start by parameterizing the breakdown energy $E_{_{\rm BD}}(t; x_i)$ according to the distance $X(t)\!\equiv\!L(t)\!-\!x_i$ between the observation point $x_i$ and the rupture edge, instead of using $t$ itself (see~Fig.~\ref{fig:fig2}). Moreover, as ordinary cracks feature localized dissipation quantified by $G_{\rm c}$, we focus on the dimensionless excess breakdown energy, defined as $\Delta{\cal E}_{_{\rm BD}}(X; x_i) \!\equiv\!(E_{_{\rm BD}}(X; x_i)-G_{\rm c})/G_{\rm c}$, where the term ``excess'' refers here to the dissipation on top of the effective fracture energy $G_{\rm c}$. To calculate $\Delta{\cal E}_{_{\rm BD}}$, consider a frictional rupture front steadily propagating at a constant velocity $c_{\rm r}$, for which the slip displacement increment at any point on the fault/interface takes the form $d\delta\=v(X; c_{\rm r},L)dX/c_{\rm r}$ (unsteady rupture propagation will be discussed below). With this relation at hand, one can use the definition of $E_{_{\rm BD}}$ to define $\Delta{\cal E}_{_{\rm BD}}$ through the following spatial integral
\begin{align}
\label{eq:DE}
&\Delta{\cal E}_{_{\rm BD}}(X; c_{\rm r}, L, \ell)=\\
&(G_{\rm c}\,c_{\rm r})^{-1}\!\!\int_\ell^X\!\left[\tau(X';c_{\rm r},L)-\tau_{\rm res} \right] v(X';c_{\rm r},L)\,dX' \nonumber\ ,
\end{align}
for $\ell\!\le\!X\!\le\!L$ (cf.~Fig.~\ref{fig:fig2}, where $\ell$, $X$ and $L$ are illustrated), where we used the fact that the integral over $0\!\le\!X\!<\!\ell$ equals $G_{\rm c}$.

Consider then a frictional interface that is described by a generic rate-and-state dependent constitutive relation~\cite{Dieterich79,Ruina1983,Rice1983,Dieterich1994b,Marone1998,Nakatani2001,Rice2001,Baumberger2006}, characterized by an $N$-shaped steady-state friction curve $\tau_{\rm ss}(v)$~\cite{Bar-Sinai2012,Bar-Sinai2014} and a single structural state field $\phi(x,t)$~\cite{Dieterich79,Ruina1983,Rice1983,Dieterich1994b,Marone1998,Nakatani2001,Rice2001,Baumberger2006,Bar-Sinai2012,Bar-Sinai2014,Dietrich2007,Nagata2012,Bhattacharya2014}, as detailed in~\cite{PartI,Barras2020} and in the Methods. For a broad range of materials, $\tau_{\rm ss}(v)$ is characterized by a non-monotonic and rather weak logarithmic rate dependence~\cite{Bar-Sinai2014}. It is well established that generic rate-and-state frictional interfaces host propagating rupture once the condition for rupture nucleation are met~\cite{Ben-Zion2001}. Suppose then that rupture nucleates at $x\=0$ (the hypocenter, cf.~Fig.~\ref{fig:fig2}) at time $t\=0$, giving rise to two symmetrically propagating frictional rupture fronts. In Fig.~\ref{fig:fig2}, we present the frictional stress $\tau(X; c_{\rm r},L)$ and slip velocity $v(X; c_{\rm r},L)$ fields of the right-propagating front at a later time $t$, as obtained by recent 2D anti-plane simulations of rate-and-state frictional interfaces~\cite{Barras2020}. In principle, the fields $\tau(X; c_{\rm r},L)$ and $v(X; c_{\rm r},L)$ can be extracted from such a simulation and plugged into Eq.~\eqref{eq:DE}. Then the integral can be evaluated numerically to yield $\Delta{\cal E}_{_{\rm BD}}$. Our goal, though, is to calculate $\Delta{\cal E}_{_{\rm BD}}$ analytically in order to gain insight into the underlying physics and then to test the resulting predictions against the simulational data.

The starting point for our development is the idea that for rate-and-state frictional interfaces we have $[\tau(X; c_{\rm r},L)-\tau_{\mbox{\scriptsize res}}]/\tau_{\rm res}\!\ll\!1$ for $X\!>\!\ell$, as is indeed observed in~Fig.~\ref{fig:fig2}. A quantitative criterion for this condition to hold is derived below. Had it been $\tau(X\!>\!\ell;c_{\rm r},L)\=\tau_{\mbox{\scriptsize res}}$, we would have $\Delta{\cal E}_{_{\rm BD}}\=0$ and the conventional slip velocity singularity $v(X; c_{\rm r},L)\!\sim\!1/\sqrt{X}$ would have been exact for $\ell\!\ll\!X\!\ll\!L$ (as illustrated in Fig.~\ref{fig:fig1} for ordinary cracks). Therefore, we treat the latter as a leading order solution and aim at expressing $\tau(X; c_{\rm r},L)-\tau_{\mbox{\scriptsize res}}$ in terms of $v(X; c_{\rm r},L)$. We then assume that the evolution of the internal state field $\phi(X,t)$ is ``fast'', i.e.~that it quickly equilibrates with $v(X; c_{\rm r},L)$. Under these conditions, we are left with $\tau(X; c_{\rm r},L)\=\tau_{\rm ss}[v(X; c_{\rm r},L)]$, where the
latter is a nonlinear relation. To allow for an analytic treatment, we further assume that the smallness of $(\tau_{\rm ss}(v)-\tau_{\rm res})/\tau_{\rm res}$ also implies the
smallness of $(v-v_{\rm res})/v_{\rm res}$, presumably justifying a linearization of $\tau_{\rm ss}(v)-\tau_{\rm res}$ around $v\=v_{\rm res}$ for the entire range $X\!>\!\ell$.

With these ideas and assumptions in mind, we obtain the following expansion
\begin{equation}
\label{eq:viscous}
\tau_{\rm ss}(v)-\tau_{\rm res} \simeq \left(d\tau_{\rm ss}(v_{\rm res})/dv\right)(v-v_{\rm res}) \approx \eta\,v \ ,
\end{equation}
where $\eta\!\equiv\!d\tau_{\rm ss}(v_{\rm res})/dv$ is an effective viscous-friction coefficient and $\tau_{\rm res}\!\gg\!v_{\rm res}d\tau_{\rm ss}(v_{\rm res})/dv$, which is typically satisfied, has been assumed. As will be shown next, this effective linear viscous-friction relation allows to gain deep analytical insight into the physics of the problem at hand~\cite{Brener2002}. Plugging Eq.~\eqref{eq:viscous} into Eq.~\eqref{eq:DE}, we obtain $\Delta{\cal E}_{_{\rm BD}}(X; c_{\rm r}, L, \ell)\=\frac{\eta}{G_{\rm c}c_{\rm r}}\!\int_\ell^X\!\left[v(X';c_{\rm r},L)\right]^2dX'$. Using then the conventional singular slip velocity field $v(X; c_{\rm r},L)\!\simeq\!2c_{\rm r}K/[\mu\,\alpha_s(c_{\rm r})\sqrt{2\pi X}]$ for anti-plane conditions~\cite{Freund1998,Broberg1999}, where $\alpha_s(c_{\rm r})\=\sqrt{1-c_{\rm r}^2/c_s^2}$ is the relativistic Lorentz factor and $K$ is the stress intensity factor, we can perform the integration to obtain
\begin{equation}
\label{eq:DE_result}
\Delta{\cal E}_{_{\rm BD}}(X; c_{\rm r}, L, \ell) \,\simeq\, \Delta\xi(c_{\rm r}) \ln(X/\ell) \ ,
\end{equation}
which is expected to hold for $\ell\!\ll\!X\!\ll\!L$, and where
\begin{equation}
\label{eq:delta_xi}
\Delta\xi(c_{\rm r}) \equiv \frac{4\,\eta \,c_{\rm r}}{\pi\,\mu\,\alpha_s(c_{\rm r})} \ .
\end{equation}
In deriving Eq.~\eqref{eq:DE_result}, we used the edge-localized energy balance $G\=K^2/[2\,\mu\,\alpha_s(c_{\rm r})]\= G_{\rm c}$~\cite{Freund1998,Broberg1999}, which is associated with dissipation on the
scale $X\!\sim\!\ell$.

The effective viscous-friction coefficient $\eta$ is positive for the $N$-shaped steady-state friction curve $\tau_{\rm ss}(v)$ because $v_{\rm res}$ typically resides on the
velocity-strengthening branch of the friction law above its minimum, $d\tau_{\rm ss}(v_{\rm res})/dv\!>\!0$. While there is ample evidence that the $N$-shaped steady-state curve is a generic property of frictional interfaces~\cite{Bar-Sinai2014}, hence $\eta\!>\!0$, it is important to note that having $\eta\=d\tau_{\rm ss}(v_{\rm res})/dv\!<\!0$ does not violate any law of nature. The point is that $\Delta{\cal E}_{_{\rm BD}}\!\times\!G_{\rm c}$ is not the total dissipation, which includes also $G_{\rm c}$, the background frictional dissipation associated with sliding against the residual stress $\tau_{\rm res}$ and radiated energy. Together, these ensure positive total dissipation and in principle one can have $\eta\!<\!0$, which implies $\Delta{\cal E}_{_{\rm BD}}\!<\!0$. This would be the case if $v_{\rm res}$ resides on a velocity-weakening branch of the friction curve, $d\tau_{\rm ss}(v_{\rm res})/dv\!<\!0$.

The excess breakdown energy $\Delta{\cal E}_{_{\rm BD}}$ in Eq.~\eqref{eq:DE_result}, which constitutes one of our major results, shows that whenever friction is rate dependent, i.e.~$\eta\!\propto\!d\tau_{\rm ss}/dv\!\ne\!0$, the breakdown energy ${\cal E}_{_{\rm BD}}$ deviates from the fracture energy $G_{\rm c}$ ($\Delta{\cal E}_{_{\rm BD}}\!\ne\!0$),  dissipation-related scale separation breaks down and $\Delta{\cal E}_{_{\rm BD}}$ explicitly depends on a macroscopic scale $X$. As $X$ is generally orders of magnitude larger than $\ell$, $\Delta{\cal E}_{_{\rm BD}}$ can in general be significantly larger than $G_{\rm c}$ (the limiting/stauration level of $\Delta{\cal E}_{_{\rm BD}}$ will be discussed below). Our next goal is to understand the range of validity of the result in Eq.~\eqref{eq:DE_result} and its relation to deviations from the conventional singularity of LEFM.\\

\noindent{\bf Relation to unconventional singularities.} The results in Eqs.~\eqref{eq:DE_result}-\eqref{eq:delta_xi} were obtained by assuming that the conventional slip velocity singularity $v(X)\!\sim\!1/\sqrt{X}$ can be treated as the leading order solution of the frictional rupture problem (behind the propagating rupture front). That is, it was implicitly assumed that in some sense the singularity order of rate-and-state frictional rupture only slightly deviates from the conventional $-\tfrac{1}{2}$ singularity. Yet, it remains unclear at this stage how Eqs.~\eqref{eq:DE_result}-\eqref{eq:delta_xi} are related to the singularity order, and in particular how these are related to the smallness of the deviation from the conventional singularity and to $[\tau(X)-\tau_{\mbox{\scriptsize res}}]/\tau_{\rm res}\!\ll\!1$.

The key to answering these questions is $\Delta\xi(c_{\rm r})$ in Eqs.~\eqref{eq:DE_result}-\eqref{eq:delta_xi} and its physical meaning. While it is common to assume that the
conventional square root singularity of LEFM remains approximately valid for frictional rupture, and while this assumption is a posteriori supported by some observations (see, for example,~\cite{Svetlizky2019, Barras2020}), for the effective
linear viscous-friction in Eq.~\eqref{eq:viscous} nothing should be assumed, the singularity order can be explicitly derived in light of the linearity of the problem. That is, we have
\begin{equation}
\label{eq:v_unconventional}
v(X; c_{\rm r})\sim \left(X/\ell\right)^{\xi(c_{\rm r})} \ ,
\end{equation}
where $\xi(c_{\rm r})$ does not necessarily and a priori equal $-\tfrac{1}{2}$, i.e.~it may correspond to an unconventional singularity emerging from the intrinsic
rate dependence of the frictional stress~\cite{Brener2002}.

Using Eq.~\eqref{eq:viscous} and specializing here for anti-plane conditions, one can show that $\xi(c_{\rm r})$ satisfies $\cot[\pi\,\xi(c_{\rm r})]\=-2\,\eta\,c_{\rm
r}/[\mu\,\alpha_s(c_{\rm r})]$ (see Methods). This relation shows that the singularity order is not a constant, but rather a dynamic quantity that varies with the rupture velocity $c_{\rm r}$.
Moreover, assuming that $\xi(c_{\rm r})$ indeed deviates from $-\tfrac{1}{2}$ only slightly, we obtain
\begin{equation}
\label{eq:unconventional}
\xi(c_{\rm r}) \simeq -\tfrac{1}{2}\big[1 - \Delta\xi(c_{\rm r})\big] \ ,
\end{equation}
where surprisingly $\Delta\xi(c_{\rm r})$ is the same one given in Eq.~\eqref{eq:delta_xi}. Consequently, Eq.~\eqref{eq:DE_result} is indeed valid when $\Delta\xi(c_{\rm r})\!\ll\!1$, i.e.~when the deviation from the conventional LEFM singularity is small. Equations~\eqref{eq:v_unconventional}-\eqref{eq:unconventional}, together with Eqs.~\eqref{eq:DE_result}-\eqref{eq:delta_xi}, constitute the major results of this work.

Equation~\eqref{eq:unconventional} shows that frictional rupture is in fact characterized by an unconventional singularity, yet that the deviation from the conventional
$-\tfrac{1}{2}$ singularity is small when $\Delta\xi$ is small. According to Eq.~\eqref{eq:delta_xi}, the latter is small when the rate dependence of friction is weak, i.e.~when the properly nondimensionalized $d\tau_{\rm ss}(v_{\rm res})/dv$ is small. This is generically the case for rate-and-state frictional interfaces. In such cases, the excess breakdown energy $\Delta{\cal E}_{_{\rm BD}}$ in Eq.~\eqref{eq:DE_result}
is proportional to the very same small quantity $\Delta\xi$ of Eq.~\eqref{eq:delta_xi}, but $\Delta{\cal E}_{_{\rm BD}}$ is not necessarily small because the smallness of $\Delta\xi$ may be compensated by an accumulated spatially-extended contribution. Therefore, we identify $\Delta\xi$ as hidden small parameter in rate-and-state frictional failure dynamics.

The origin of this small parameter is the rate dependence of the frictional stress, which in turn implies that the strict scale separation assumed in LEFM is only approximately valid in frictional rupture dynamics (manifested in the slow decay of $\tau(X)$ toward $\tau_{\rm res}$, while satisfying $[\tau(X)-\tau_{\mbox{\scriptsize res}}]/\tau_{\rm res}\!\ll\!1$). Moreover, some physical quantities (e.g.~$\Delta{\cal E}_{_{\rm BD}}$) may be more strongly affected than others (e.g.~$\xi$) by the lack of strict scale separation. Finally, we note that for other interfacial constitutive relations $\Delta\xi$ may not be small and additional new physics may emerge. Such situations will not be extensively discussed here, but will be mentioned below in relation to seismological observations. But first, we set out to quantitatively test the predictions of the theory against detailed cutting-edge computer simulations~\cite{Barras2020}.\\

\noindent{\bf Testing the theory.} In order to test the theoretical predictions in Eqs.~\eqref{eq:DE_result}-\eqref{eq:unconventional}, we consider the recent computer simulations of~\cite{Barras2020}, where generic rate-and-
state dependent frictional dynamics have been studied. In these 2D anti-plane simulations, frictional rupture fronts spontaneously emerge, allowing accurate calculations of all of the physically relevant quantities discussed above. In particular, it has been shown that the near rupture edge fields (e.g.~those shown in Fig.~\ref{fig:fig2}) follow the conventional LEFM $-\tfrac{1}{2}$ singularity to a very good approximation~\cite{Barras2020}. That is, while $\Delta\xi$ in Eq.~\eqref{eq:unconventional} has not been explicitly calculated, the results of~\cite{Barras2020} clearly indicate that $\Delta\xi\!\ll\!1$, which is precisely the validity condition of Eqs.~\eqref{eq:DE_result}-\eqref{eq:delta_xi}.

\begin{figure*}[ht!]
\centering
\includegraphics[width=0.87\textwidth]{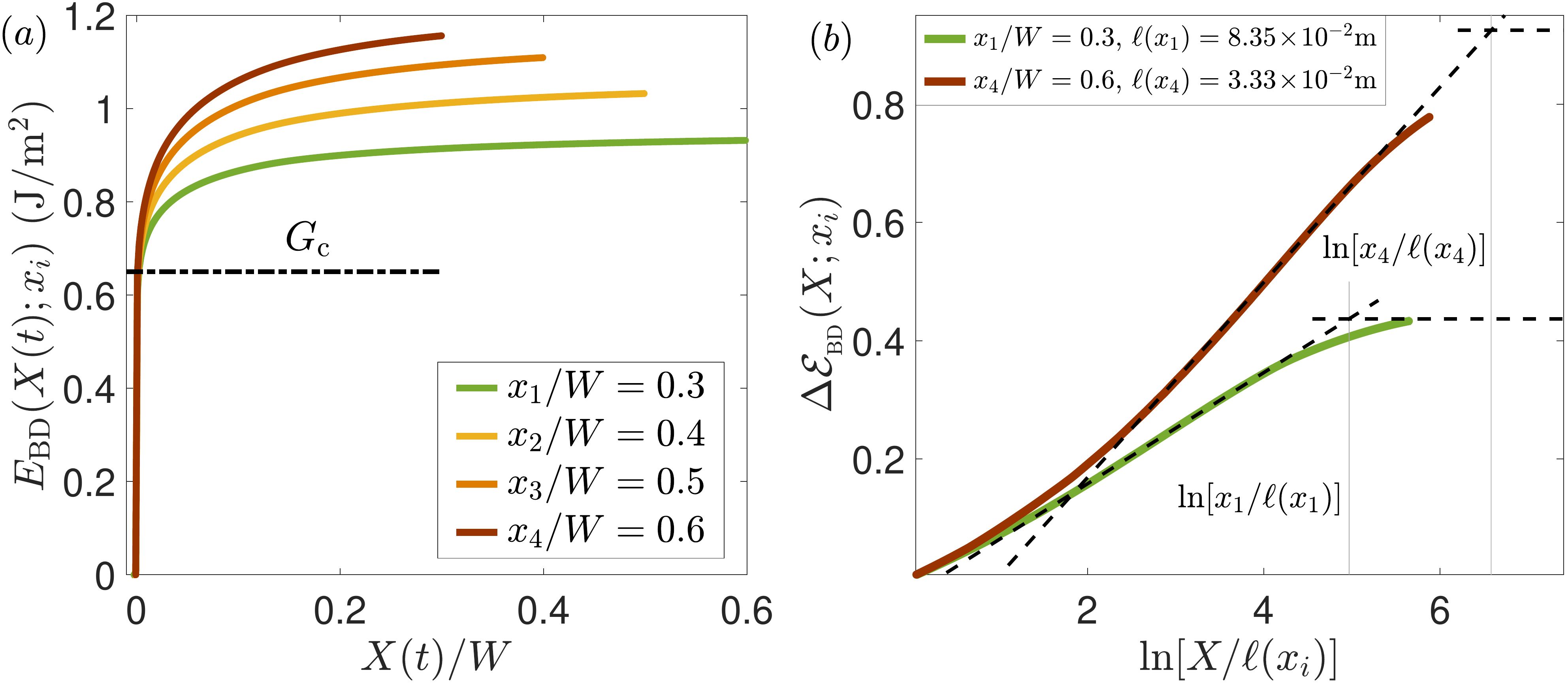}
\caption{{\bf The breakdown energy of frictional rupture}. (a) The breakdown energy $E_{_{\rm BD}}(X(t); x_i)$ as a function of $X(t)/W$ for $t\!>\!t_{x_i}$ (see text for definition), obtained in numerical simulations of rate-and-state frictional interfaces~\cite{Barras2020}, for $4$ different observation points $x_{1-4}$ (see legend). $W$ is the fault/interface half-length and we set $t_{x_i}\!=\!0$ for $i\!=\!1\!-\!4$ (for presentational convenience). All curves perfectly overlap over a short lengthscale, which identifies with $\ell$ (see Fig.~\ref{fig:fig2}), defining the effective fracture energy $G_{\rm c}$ (dashed-dotted horizontal line), but branch out on larger scales. See text for additional discussion. (b) $\Delta{\cal E}_{_{\rm BD}}(X; x_i)$, corresponding to the data presented in panel a, vs.~$\ln[X/\ell(x_i)]$ for $x_1/W\!=\!0.3$ and $x_4/W\!=\!0.6$ (see legend). $\Delta{\cal E}_{_{\rm BD}}(X; x_i)$ follows a logarithmic behavior at an intermediate range, as highlighted by the titled dashed lines (the slope of the lower line is $0.094$ and that of the upper one is $0.165$, their ratio is $1.76$). $\Delta{\cal E}_{_{\rm BD}}(X; x_i)$ presumably crosses over, roughly at $\ln[x_i/\ell(x_i)]$ (light gray vertical lines), to a plateau (illustrated by the horizontal dashed lines).}
\label{fig:fig3}
\end{figure*}

In Fig.~\ref{fig:fig3}a, we present $E_{_{\rm BD}}(X(t); x_i)$ for $4$ different observation points $x_{1-4}$ along the fault/interface. It is observed that the $E_{_{\rm BD}}(X(t); x_i)$ curves for all $x_i$'s overlap on a small lengthscale and then branch out. The curves then keep on increasing and appear to saturate at $x_i$-dependent values that are substantially larger than the value of $E_{_{\rm BD}}$ at the branching out point. This behavior is qualitatively different from the one of ordinary cracks, cf.~Fig.~\ref{fig:fig1}. On the other hand, it appears to be consistent with the theoretical predictions of Eq.~\eqref{eq:DE_result}, not considering for the meantime the $x_i$ dependence and the saturation (to be discussed later).

A clear signature of the analytic prediction in Eq.~\eqref{eq:DE_result} is the logarithmic dependence of $\Delta{\cal E}_{_{\rm BD}}$ on $X$ in the intermediate range $\ell\!\ll\!X\!\ll\!L$. To test this prediction, we need to identify $G_{\rm c}$, which is nothing but the value of $E_{_{\rm BD}}$ at the branching out point. Indeed, it was shown in~\cite{Barras2020} that this value of $G_{\rm c}$ is exactly the one that balances the elastic energy flux $G$ into the edge region, as determined from the extracted stress intensity factor $K$. Consequently, the LEFM edge-localized energy balance $G\!\approx\!G_{\rm c}$ is satisfied to a very good approximation, which in turn determines the rupture velocity $c_{\rm r}$~\cite{Barras2020}. Moreover, $G_{\rm c}$ allows to explicitly extract the localization length $\ell$. Having at hand both $G_{\rm c}$ and $\ell$, we plot in Fig.~\ref{fig:fig3}b $\Delta{\cal E}_{_{\rm BD}}$ (corresponding to the data of Fig.~\ref{fig:fig3}a) against $\ln(X/\ell)$, for the lowest and largest $x_i$'s. It is observed that $\Delta{\cal E}_{_{\rm BD}}$  depends logarithmically on $X$ in an intermediate range (for the two extreme values of $x_i$), as predicted analytically in Eq.~\eqref{eq:DE_result}, lending strong support to the theory.

The slope/prefactor of the logarithmic relation depends on the observation point $x_i$, which in turn implies that $\Delta\xi$ in Eqs.~\eqref{eq:DE_result}-\eqref{eq:delta_xi} depends on $x_i$. Equation~\eqref{eq:delta_xi} predicts, assuming that the effective linear viscous-friction coefficient $\eta$ is independent of $x_i$, that the observed $x_i$ dependence is attributed to the rupture propagation velocity $c_{\rm r}$, in particular to the combination $c_{\rm r}/\alpha_s(c_{\rm r})$. Indeed, frictional rupture in the numerical simulation corresponding to Fig.~\ref{fig:fig3} continuously accelerated~\cite{Barras2020}, i.e.~$\dot{c}_{\rm r}(t)\!>\!0$, where  $c_{\rm r}(t_{x_1})\=0.94c_s$ and $c_{\rm r}(t_{x_4})\=0.983c_s$. Using these values inside Eq.~\eqref{eq:delta_xi}, the theory predicts the ratio of the slopes in Fig.~\ref{fig:fig3}b to be $1.94$. This prediction is in reasonably good quantitative agreement with the observed ratio, which equals $1.76$. Finally, the individual slopes satisfy $\Delta\xi(c_{\rm r})\!\sim\!{\cal O}(10^{-1})$, which is in agreement with a direct estimation of $\Delta\xi(c_{\rm r})$ according to Eq.~\eqref{eq:delta_xi}, using the constitutive parameters of the numerical simulations~\cite{Barras2020}.

Taken together, these results provide direct support to the theoretical predictions. In particular, the results show that $\Delta{\cal E}_{_{\rm BD}}$ can be, and in fact is, quite significantly larger than $\Delta\xi(c_{\rm r})\!\ll\!1$. This happens due to accumulated spatial contribution associated with the huge difference between $X$ --- that can reach the fault/interface size --- and the localization length $\ell$ (and
despite the logarithmic dependence on their ratio). We thus conclude that for rate-and-state frictional interfaces, the non-edge-localized excess breakdown energy in Eq.~\eqref{eq:DE_result} is a product of a typically small number, given by Eq.~\eqref{eq:delta_xi}, and an accumulated spatially-extended contribution that can compensate the smallness of $\Delta\xi(c_{\rm r})$. Consequently, the breakdown energy $E_{_{\rm BD}}$ can in general deviate significantly from the fracture energy $G_{\rm c}$.\\

\noindent{\bf The position dependence of the breakdown energy and its saturation level.} How large can the deviation of $E_{_{\rm BD}}$ from $G_{\rm c}$ be? What determines the magnitude of the deviation? In the example shown in Fig.~\ref{fig:fig3}a, the deviation can be as large as $\sim\!100\%$, but more importantly it is observed that the saturation value of $E_{_{\rm BD}}(X; x_i)$ depends on the observation point $x_i$. That is, in addition to the $x_i$ dependence discussed above in relation to non-steady rupture propagation, the simulational results indicate an intrinsic relation between the observation point and the saturation value of $E_{_{\rm BD}}$. It is clear that the $\ln(X/\ell)$ dependence of $\Delta{\cal E}_{_{\rm BD}}$ in Eq.~\eqref{eq:DE_result}, which was discussed and validated in the intermediate range $\ell\!\ll\!X\!\ll\!L$ in Fig.~\ref{fig:fig3}b, cannot persist indefinitely. This is simply the case because the logarithmic dependence is directly related to the singular part of $v(X)$, which is no longer dominant at large $X$.

To understand the behavior of $\Delta{\cal E}_{_{\rm BD}}(X)$ at large $X$, note that the during crack propagation the relation $L(t)\=x_i+X(t)$ holds for $L(t)\!\ge\!x_i$, where both $L(t)$ and $X(t)$ increase, while $x_i$ is fixed. At short propagation times, measured relative to the time at which $L(t)\=x_i$, we have $X\!\ll\!x_i$. At intermediate propagation times, $\Delta{\cal E}_{_{\rm BD}}(X; x_i)$ varies logarithmically with $X$, as demonstrated in Fig.~\ref{fig:fig3}b, and finally, at long propagation times, we have $X(t)\!\to\!L(t)$, which implies $X\!\gg\!x_i$ and for which the logarithmic law is not valid anymore. If $\tau(X)$ approaches $\tau_{\rm res}$ for large $X$ in a way that leads to the saturation of $\Delta{\cal E}_{_{\rm BD}}(X)$, then we expect the logarithmic behavior of $\Delta{\cal E}_{_{\rm BD}}(X; x_i)$ to cross over to a plateau on a scale $X\!\sim\!x_i$, i.e.~when $\ln(X/\ell)$ roughly equals $\ln(x_i/\ell(x_i)]$. This prediction is tested and supported in Fig.~\ref{fig:fig3}b, demonstrating that $\Delta{\cal E}_{_{\rm BD}}(X; x_i)$ indeed crosses over to a plateau on a scale $X\!\sim\!x_i$. This is a surprising and somewhat non-intuitive result that shows that it is not the rupture size $L$ per se that determines the magnitude of the deviation of $E_{_{\rm BD}}$ from $G_{\rm c}$, but rather the observation point $x_i$. Obviously, larger ruptures generally allow
larger $x_i$, so in general these can feature larger deviations of $E_{_{\rm BD}}(X; x_i)$ from $G_{\rm c}$.

This physical insight can be used to quantitatively predict the $E_{_{\rm BD}}(X; x_i)$ curves, shown in Fig~\ref{fig:fig3}a, over the full range of $X$'s and different $x_i$'s. To that aim, we need to include higher order, non-singular contributions to $v(X; c_{\rm r},L)$. This is done by using Broberg's full-field solution for a self-similar crack propagating at a constant velocity $c_{\rm r}$, which takes the form $v(X; c_{\rm r},L)\=c_{\rm r}\sqrt{\frac{8\,G_c\,L}{\pi\,\mu\,\alpha_s(c_{\rm r})}}\big/\!\sqrt{L^2-(L-X)^2}$~\cite{Broberg1999}, and is valid for $\ell\!\le\!X\!\le\!L$. Furthermore, as the full-field expression anyway requires numerical integration in Eq.~\eqref{eq:DE}, we can relax the assumption that $\tau_{\rm ss}(v)-\tau_{\rm res}$ can be linearized around $v_{\rm res}$. While this assumption appears plausible, the smallness of $(\tau_{\rm ss}(v)-\tau_{\rm res})/\tau_{\rm res}$ does not strictly guarantee that the linear relation in Eq.~\eqref{eq:viscous} is quantitatively accurate over the entire range $X\!>\!\ell$, which involves a rather broad range of slip velocities. Consequently, we use the fully nonlinear $N$-shaped $\tau_{\rm ss}(v)$ (see Methods) instead of the linearized one of Eq.~\eqref{eq:viscous}, and use Broberg's full-field solution $v(X; c_{\rm r},L)]$ for $v$, when numerically evaluating the integral in Eq.~\eqref{eq:DE}. It is important to note that as Broberg's full-field solution is obtained in the framework of LEFM, the proposed procedure is still perturbative in nature; that is, it employs an LEFM solution as the leading order contribution in order to calculate the relevant dissipation integral in the main approximation~\cite{Brener2002}.
\begin{figure}[ht!]
\centering
\includegraphics[width=0.45\textwidth]{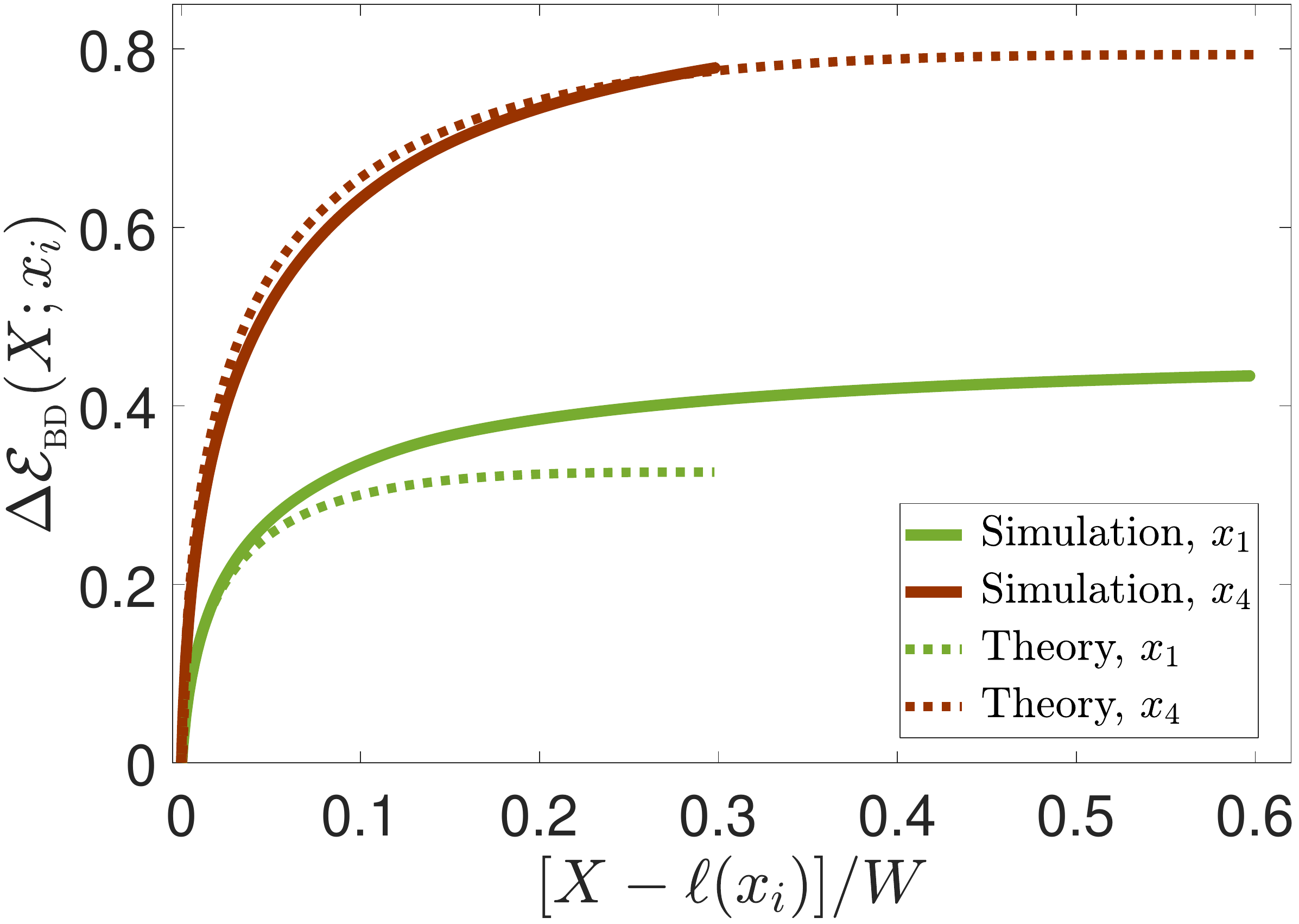}
\caption{{\bf Testing the theoretical predictions}. Comparison of the simulational results for $\Delta{\cal E}_{_{\rm BD}}(X; x_i)$ (solid lines, see legend), presented earlier in Fig.~\ref{fig:fig3}a, and the theoretical prediction $\Delta{\cal E}_{_{\rm BD}}(X; c_{\rm r}(x_i), L\!=\!x_i, \ell(x_i))$ (dashed lines, see legend) obtained using Eq.~\eqref{eq:DE} (see text for additional details), vs.~$[X-\ell(x_i)]/W$ for $x_1/W\!=\!0.3$ and $x_4/W\!=\!0.6$ (see legend). The values of $\ell(x_i)$ are given in Fig.~\ref{fig:fig3}b. Note that the theoretical curves, by construction, extend up to $X\!=\!x_i$ (recall that $\ell(x_i)\!\ll\!x_i$) and that the simulational ones are limited by the fault/interface half-length $W$, the smaller $x_i$ the larger the maximal $X$.}
\label{fig:fig4}
\end{figure}

$\Delta{\cal E}_{_{\rm BD}}(X; c_{\rm r}, L, \ell)$ of Eq.~\eqref{eq:DE} is a spatial integral over a snapshot of the rupture fields, which in itself is independent of the observation point $x_i$. The observation point dependence is introduced in two steps, corresponding to difference pieces of physics. First, as the integral in Eq.~\eqref{eq:DE} extends up to $X\=L$, the insight about the saturation at $X\!\sim\!x_i$ can be captured by setting $L\=x_i$. This saturation is totally unrelated to the additional $x_i$ dependence introduced by the non-steadiness of rupture propagation. The latter, as already discussed earlier, is captured by setting $c_{\rm r}\=c_{\rm r}(x_i)$ and $\ell\=\ell(x_i)$. Consequently, we calculated $\Delta{\cal E}_{_{\rm BD}}(X; c_{\rm r}(x_i), L\=x_i, \ell(x_i))$ of Eq.~\eqref{eq:DE} using the fully nonlinear $N$-shaped $\tau_{\rm ss}(v)$ (see Methods) and Broberg's full-field solution $v(X; c_{\rm r}(x_i),L\=x_i)$, and compared it to the simulational results. The comparison, which is presented in Fig.~\ref{fig:fig4}, reveals reasonable quantitative agreement between the theory and the simulations, lending strong support to former.\\

\noindent{\bf Possible implications for seismological observations.} As explained above, the breakdown energy constitutes an important contribution to the total frictional rupture energy budget, which includes also the background frictional dissipation (heat) and the radiated energy. In the context of earthquake physics, these three contributions sum up to the potential energy release during an earthquake~\cite{Kanamori2000}. It would be interesting to discuss whether, and if so to what extent, our theory might have some implications for seismological observations.
The latter typically aim at using source spectra to obtain coarse-grained average estimates of the following quantity~\cite{Palmer1973,Abercrombie2005,Tinti2005,Viesca2015,Nielsen2016,Brantut2017}
\begin{equation}
\label{eq:Gf}
G_{\rm f}(\delta)\equiv \int_0^{\delta}[\tau(\delta')-\tau(\delta)]\,d\delta' \ .
\end{equation}
Note that $G_{\rm f}(\delta)$ differs from the breakdown energy $E_{_{\rm BD}}(\delta; x_i)\=\int_0^{\delta}[\tau(\delta'; x_i)\!-\!\tau_{\rm res}]\,d\delta'$ in two respects. First, it makes no reference to a fault observation point $x_i$. Second, the reference stress used in it is $\tau(\delta)$, rather than the constant residual stress $\tau_{\rm res}$.

Before discussing seismological observations, let us calculate $G_{\rm f}(\delta)$ in the framework of the theory developed in this work. As explained above, the dissipation in the spatial range $0\!\le\!X\!\le\!\ell$ near the rupture edge gives rise to a well-defined effective fracture energy $G_{\rm c}$, marked in Fig.~\ref{fig:fig3}a. The edge-localized dissipation $G_{\rm c}$ is related to a strong strength reduction (cf.~Fig.~\ref{fig:fig2}) over a characteristic slip displacement $\delta_c$, such that $G_{\rm f}(\delta_{\rm c})\=G_{\rm c}$ (note that $G_{\rm c}$ of Fig.~\ref{fig:fig3}a, which is based on $E_{_{\rm BD}}$, slightly differs in its value from the one associated with $G_{\rm f}$ due to the difference in the definition of these quantities). This strong frictional strength reduction is associated in the rate-and-state constitutive framework with the evolution of the internal state field $\phi$. It has been shown~\cite{Cocco2002,Bizzarri2003} that while the rate-and-state constitutive framework does not make explicit reference to $\delta$, the strength reduction from $\tau_0$ --- reached after the very initial increase in slip velocity near the rupture edge --- to $\tau_{\rm c}$ at $\delta\=\delta_{\rm c}$ (where $\tau_{\rm c}$ is close to, but still larger than, $\tau_{\rm res}$) follows an effective linear slip-weakening law of the form $\tau(\delta)\!\simeq\!\tau_0-(\tau_0-\tau_{\rm c})\delta/\delta_{\rm c}$. Plugging the latter into Eq.~\eqref{eq:Gf}, we obtain
\begin{equation}
\label{eq:Gf_small_delta}
G_{\rm f}(\delta) \sim \delta^2 \qquad \hbox{for} \qquad \delta\!\le\!\delta_{\rm c} \ .
\end{equation}

According to Eq.~\eqref{eq:Gf_small_delta}, $G_{\rm f}(\delta)$ follows a quadratic power law for $\delta\!\le\!\delta_{\rm c}$, i.e.~for $G_{\rm f}(\delta)\!\le\!G_{\rm c}$. For $\delta\!>\delta_{\rm c}$, where the frictional strength slowly reduces from $\tau_{\rm c}$ to the residual stress $\tau_{\rm res}$ (cf.~Fig.~\ref{fig:fig2}, $\tau_{\rm c}$ is not marked), $G_{\rm f}(\delta)$ is associated with the dissipation in the extended spatial range $X\!>\!\ell$. Our viscous-friction theory predicts that the excess dissipation in this range --- on top of $G_{\rm c}$ --- is intimately related to the emergence of unconventional singularities in frictional rupture, which in turn mainly depend on the rate-dependence of friction (and not on the internal state field $\phi$). To obtain $G_{\rm f}(\delta)$ in this regime, we use the slip velocity in Eq.~\eqref{eq:v_unconventional}, the viscous-friction relation in Eq.~\eqref{eq:viscous} and the steady-state relation $v\!=\!c_{\rm r}\,d\delta/dX$. These yield $\tau(\delta)\!-\!\tau_{\rm res}\!\sim\!\delta^{\frac{\xi}{1+\xi}}$, which upon substitution inside Eq.~\eqref{eq:Gf} leads to
\begin{equation}
\label{eq:Gf_large_delta}
G_{\rm f}(\delta)-G_{\rm c} \sim \delta^{\frac{1+2\xi}{1+\xi}} \qquad \hbox{for} \qquad \delta\!>\!\delta_{\rm c} \ .
\end{equation}
For the conventional singularity, $\xi\=-\tfrac{1}{2}$, Eq.~\eqref{eq:Gf_large_delta} predicts that $G_{\rm f}$ is independent of $\delta$ for $\delta\!>\!\delta_{\rm c}$, as expected (in fact, the pre-factor also vanishes in this case, implying $G_{\rm f}\=G_{\rm c}$). In cases in which unconventional singularities emerge, Eq.~\eqref{eq:Gf_large_delta} predicts a power law that depends on the unconventional singularity order $\xi$.
\begin{figure}[ht!]
\centering
\includegraphics[width=0.47\textwidth]{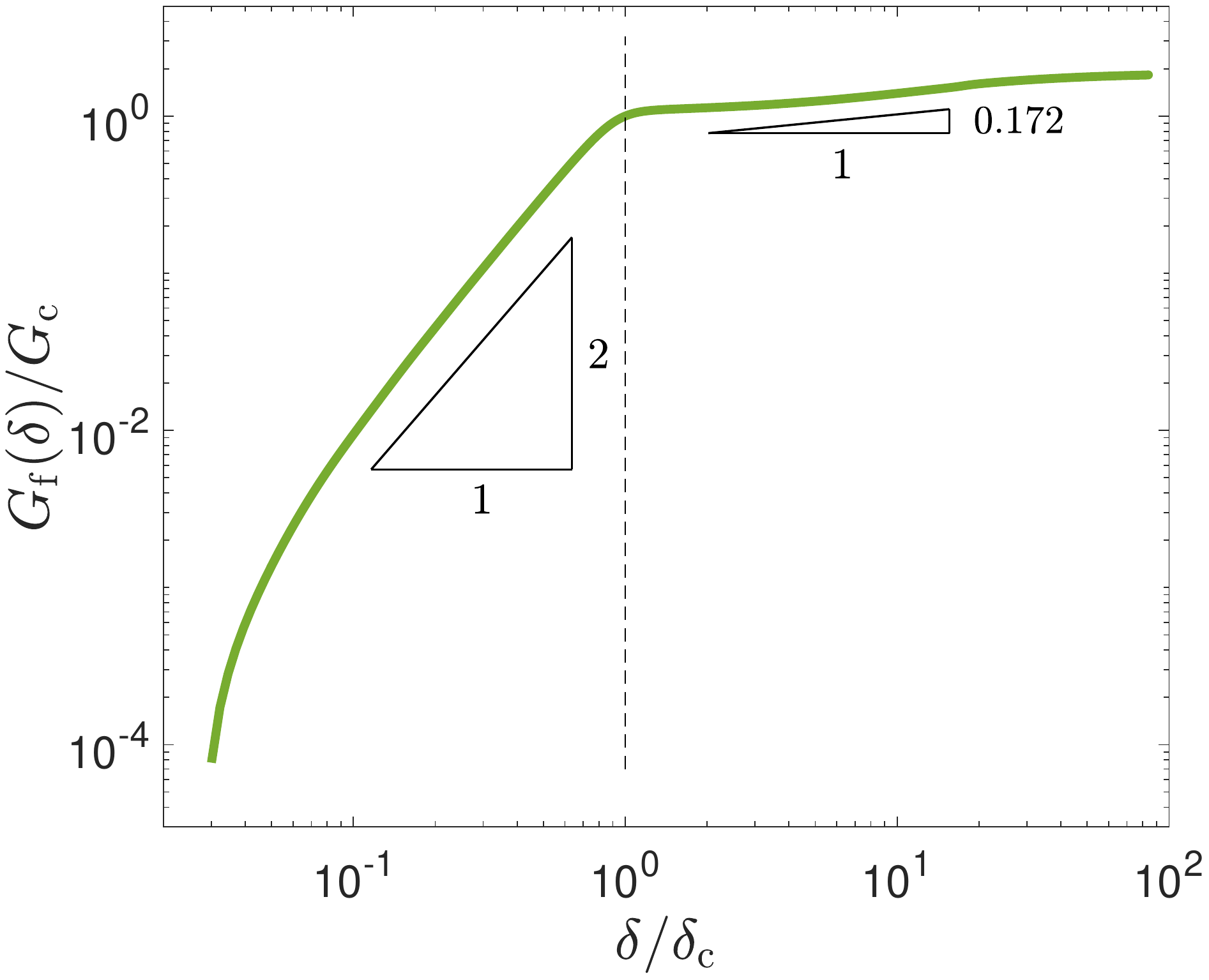}
\caption{{\bf The slip dependence of rupture-related dissipation}. $G_{\rm f}(\delta)$, defined in Eq.~\eqref{eq:Gf}, for the very same rupture simulation whose results are presented in Figs.~\ref{fig:fig3}-\ref{fig:fig4} (for the observation point $x_1/W\!=\!0.3$). $G_{\rm f}(\delta)$ features two power laws, marked by the two triangles, in agreement with the theoretical predictions in Eqs.~\eqref{eq:Gf_small_delta}-\eqref{eq:Gf_large_delta}. The crossover between the two power laws, marked by the vertical dashed line, occurs at $(\delta_{\rm c}, G_{\rm c})$, as predicted theoretically. $G_{\rm c}$ and $\delta_{\rm c}$ are used to normalize $G_{\rm f}$ and $\delta$, respectively (the value used for the former, $G_{\rm c}\!=\!0.51$J/m$^2$, is slightly smaller than the one observed in Fig.~\ref{fig:fig3}a due to the difference in the definition of $G_{\rm f}$ and $E_{_{\rm BD}}$). Finally, comparing the power law exponent in the $\delta\!>\!\delta_{\rm c}$ regime, $0.172$, to the analytic prediction in Eq.~\eqref{eq:Gf_large_delta} and using the transformation $\xi\!=\!-\tfrac{1}{2}(1\!-\!\Delta\xi)$ (cf.~Eq.~\eqref{eq:unconventional}), one obtains $\Delta\xi\!=\!0.094$. The latter is in perfect quantitative agreement with $\Delta\xi$ extracted in Fig.~\ref{fig:fig3}a (see also Eqs.~\eqref{eq:DE_result}-\eqref{eq:delta_xi}).}
\label{fig:fig5}
\end{figure}

Our theory thus predicts that $G_{\rm f}(\delta)$ follows a quadratic power law for $\delta\!\le\!\delta_{\rm c}$, cf.~Eq.~\eqref{eq:Gf_small_delta}, which is associated with the strong frictional strength reduction taking place near the rupture edge. The quadratic law is a signature of an effective linear slip-weakening characterizing this strength reduction process. At $G_{\rm f}(\delta_{\rm c})\=G_{\rm c}$, the theory predicts a crossover to another power law, valid for $\delta\!>\!\delta_{\rm c}$ (cf.~Eq.~\eqref{eq:Gf_large_delta}), which is associated with spatially-extended dissipation and is determined by the unconventional singularity order $\xi$. These predictions are being tested in Fig.~\ref{fig:fig5} for the smallest $x_i$ data presented earlier in Figs.~\ref{fig:fig3}-\ref{fig:fig4} (green curves). The numerical results quantitatively agree with the theoretical predictions, revealing a quadratic power law at small $\delta$ as predicted by~Eq.~\eqref{eq:Gf_small_delta}, and a weaker power low (here with an exponent $0.172$) for $G_{\rm f}(\delta)\!>\!G_{\rm c}$, as predicted by~Eq.~\eqref{eq:Gf_large_delta}. Using the latter, together with the transformation $\xi\!=\!-\tfrac{1}{2}(1\!-\!\Delta\xi)$ (cf.~Eq.~\eqref{eq:unconventional}), one obtains $\Delta\xi\!=\!0.094$, which is in perfect quantitative agreement with $\Delta\xi$ extracted in Fig.~\ref{fig:fig3}a (see also Eqs.~\eqref{eq:DE_result}-\eqref{eq:delta_xi}). Note that $\Delta\xi$ exhibits some dependence on the observation point $x_i$ (cf.~Fig.~\ref{fig:fig3}a), which is not discussed here since seismological observations --- to be considered next --- completely lack the spatial resolution required to reveal this dependence.

As explained above, seismological observations aim at using source spectra to obtain coarse-grained average estimates of $G_{\rm f}$ in Eq.~\eqref{eq:Gf}, e.g.,~see~\cite{Abercrombie2005,Tinti2005,Viesca2015,Nielsen2016,Brantut2017}. Yet, such seismological observations completely lack the spatiotemporal resolution to probe the slip $\delta$ as a function of time at a given observation point $x_i$ on the fault. Instead, it is common to plot the seismological estimate of $G_{\rm f}$ as a function of the total average slip $\bar{\delta}$ in an earthquake, making no explicit reference to the spatiotemporal evolution of slip during rupture. Moreover, it is common to superimpose the seismological estimates of $G_{\rm f}$ vs.~$\bar{\delta}$ for many earthquakes (including both crack-like and pulse-like events) occurring on different faults in a single plot, while it is not a priori clear that the data should at all collapse on a master curve. Finally, natural faults exhibit richer constitutive behaviors at high slip velocities (e.g.~related to flash weakening and thermal pressurization~\cite{Viesca2015}) compared to the rate-and-state framework used in this work, feature geometrical complexity and are 3D in nature. Yet, with these caveats in mind and following other authors~\cite{Abercrombie2005,Tinti2005,Viesca2015,Nielsen2016}, we identify $\bar{\delta}$ with $\delta$ and discuss the qualitative salient features of the theoretical predictions in Eqs.~\eqref{eq:Gf_small_delta}-\eqref{eq:Gf_large_delta} in relation to the available $G_{\rm f}$ vs.~$\bar{\delta}$ seismological observations.

Various authors compiled seismological observations from many earthquakes on different faults, spanning a broad range of total average slip $\bar{\delta}$, ranging from the micron scale to the scale of tens of meters~\cite{Abercrombie2005,Tinti2005,Viesca2015,Nielsen2016}. Several authors~\cite{Tinti2005,Viesca2015} reported $G_{\rm f}(\bar{\delta})\!\sim\!\bar{\delta}^2$ for relatively small $\bar{\delta}$, consistently with Eq.~\eqref{eq:Gf_small_delta}, i.e.~with an effective linear slip-weakening behavior near the rupture edge. Others, e.g.~\cite{Abercrombie2005,Nielsen2016}, suggested a weaker-than-quadratic small $\bar{\delta}$ power law and interpreted it in terms of a sub-linear slip-weakening behavior near the rupture edge. None of these, to the best of our knowledge, managed to single out $G_{\rm c}$ --- i.e.~the part of $G_{\rm f}$ that is balanced the edge-localized energy flux $G$ and that in turn controls the rupture propagation velocity --- from the data, as our theory allows.

Probably most relevant for our theoretical predictions are the data compiled in~\cite{Viesca2015}, where a quadratic power law $G_{\rm f}(\bar{\delta})\!\sim\!\bar{\delta}^2$ at small $\bar{\delta}$ appears to cross over to a weaker power law $G_{\rm f}(\bar{\delta})\!\sim\!\bar{\delta}^{2/3}$ at large $\bar{\delta}$. This behavior appears to be in qualitative agreement with the theoretical predictions in Eqs.~\eqref{eq:Gf_small_delta}-\eqref{eq:Gf_large_delta}, suggesting that different physics controls the two power law regimes, and in particular that the latter is associated with an unconventional singularity and a dissipation contribution from a spatially-extended region behind the rupture edge. Interestingly, the two power laws suggested in~\cite{Viesca2015} have been interpreted in terms of a thermal pressurization constitutive model, where the fluid pore pressure plays a central role. The quadratic power law has been interpreted to correspond to an effective linear slip-weakening behavior associated with undrained conditions and the $\tfrac{2}{3}$ power law with drained conditions~\cite{Viesca2015}. Most interestingly, the $\tfrac{2}{3}$ power law regime has been related to an unconventional singularity of order $\xi\=-\tfrac{1}{4}$, associated with the thermal pressurization model under drained conditions and corresponding to large slips accumulated far behind the rupture edge. In fact, substituting $\xi\=-\tfrac{1}{4}$ in Eq.~\eqref{eq:Gf_large_delta}, one obtains a $\tfrac{2}{3}$ power law, even though the linear viscous-friction approximation of Eq.~\eqref{eq:viscous} does not seem to be directly relevant to the analysis of~\cite{Viesca2015}.\\

\noindent{\bf \large Discussion}

In this work we developed a theory that elucidates the interrelation between unconventional singularities, scale separation and energy balance in frictional rupture. We have shown that the intrinsic rate dependence of friction, $d\tau_{\rm ss}(v)/dv\!\ne\!0$, generically leads to deviations from the conventional LEFM near-edge singularity. It is this rate dependence, which in turn implies that the frictional stress is self-selected, that leads to the emergence of singular fields different from those of LEFM. For the widespread rate-and-state friction constitutive law, these deviations can be small, yet they are accompanied by a non-edge-localized breakdown energy that significantly deviates from the edge-localized dissipation.

The developed theory sheds basic light on frictional rupture energy balance and the underlying lengthscales. The crux of the theory is the identification of a hidden small parameter $\Delta\xi$ that quantifies the deviation from the conventional LEFM singularity and that is intrinsically related to the rate dependence of friction. The theory quantitatively explains recent puzzling observations in cutting-edge numerical simulations and offers predictions that are amenable to laboratory testing using available techniques~\cite{Svetlizky2019}. Finally, the theory offers tools and concepts that can be used to interpret seismological estimates of earthquake breakdown energies.

The concepts and ideas developed in this work are applicable to more complicated interfacial constitutive relations, incorporating even richer multiphysics of frictional systems. These can include healing, pore fluid effects, thermal pressurization, flash heating, off-fault damage and plasticity, and more. The developed theory remains valid as long as the frictional stress continues to evolve behind the rupture front over scales much larger than the localization (cohesive/process zone) length $\ell$, implying that strict scale separation does not hold. In the most general case, the power law exponent $\xi(c_{\rm r})$ in Eq.~\eqref{eq:v_unconventional} is not necessarily close to $-\tfrac{1}{2}$ (i.e.~$\Delta\xi$ is not necessarily small). This generalized theory will be addressed elsewhere, and applications to specific interfacial constitutive relations are expected to emerge in the future.\\

\noindent{\bf \large Methods}

This work is analytic in nature and all of the derivations are detailed in the text, except for the solution for the unconventional singularity order, which is provided below. The theoretical predictions are compared to numerical results that have been published in~\cite{Barras2020} based on 2D anti-plane spectral boundary integral method simulations~\cite{Geubelle1995,Morrissey1997,Breitenfeld1998}. These numerical simulations employed a rate-and-state friction constitutive relation $\tau\=\sigma\sgn(v)f(|v|,\phi)$, where $\sigma$ is the normal stress and $f(|v|,\phi)\=[1+b\log(1+\phi/\phi_*)][f_0/\!\sqrt{1+(v_*/v)^2}+a\log(1+|v|/v_*)]$. The internal state field $\phi$ satisfies $\dot\phi\=1\!-\!\sqrt{1+(v_*/v)^2}\,|v|\phi/D$ and the values of the parameters appear in Table I in~\cite{PartI}. Under steady-state conditions, $\dot\phi\=0$, the frictional strength $\tau_{\rm ss}(v)$ follows an $N$-shaped curve, as plotted in Fig.~2a of~\cite{PartI} and in Fig.~1b of~\cite{Barras2020}, and as supported by numerous experiments~\cite{Bar-Sinai2014}. The numerical results of~\cite{Barras2020} have been presented in this work in different forms, depending on the theoretical predictions being tested, as detailed in the text.

The unconventional singularity order $\xi$ can be obtained by considering the interfacial boundary condition for 2D anti-plane steadily propagating rupture~\cite{Weertman1980}
\begin{equation}
\label{eq:boundary_integral}
\tau(X) = \frac{\mu\,\alpha_s(c_{\rm r})}{2\pi c_{\rm r}} \int_0^\infty \!\frac{v(X')}{X'-X}\,dX' \ ,
\end{equation}
where the left-hand-side is the frictional strength and the right-hand-side is the shear stress at the interface, as obtained from bulk elastodynamics~\cite{Weertman1980}. Using $\tau_{\rm ss}(v)$ of Eq.~\eqref{eq:viscous} for $\tau(X)$ and invoking the asymptotic power law ansatz in Eq.~\eqref{eq:v_unconventional}, Eq.~\eqref{eq:boundary_integral} implies that the unconventional singularity order $\xi$ satisfies $\cot(\pi\,\xi)\=-2\,\eta\,c_{\rm r}/(\mu\,\alpha_s(c_{\rm r}))$, as reported in the text. Finally, plugging into the last relation Eq.~\eqref{eq:unconventional} and expanding to the leading order in $\Delta\xi$, the latter is calculated and is shown to identify with Eq.~\eqref{eq:delta_xi}, as stated in the text.\\

\noindent{\bf \large Data availability}

The authors declare that the main data supporting the findings of this study are available within the article. Extra data are available from the corresponding author upon request.\\


\vspace{1cm}

\noindent{\bf \large Acknowledgements}

We are grateful to F.~Barras for critically reading the manuscript, for pushing us to think about seismological observations, and for his help with Figs.~\ref{fig:fig2},~\ref{fig:fig3} and~\ref{fig:fig5}. We thank T.~Roch for his help with Figs.~\ref{fig:fig2}-\ref{fig:fig3} and Y.~Lubomirsky for his help with the numerical integration in relation to Fig.~\ref{fig:fig4}. E.B.~acknowledges support from the Israel Science Foundation (Grants No.~295/16 and~1085/20), the Ben May Center for Chemical Theory and Computation, and the Harold Perlman Family.\\

\noindent{\bf \large Author contributions}

E.A.B.~and E.B.~designed the research, performed and research and wrote the manuscript.

\end{document}